\documentclass[10pt,aps,twoside,superscriptaddress,twocolumn,floatfix,a4paper,pra,longbibliography]{revtex4-2}
\usepackage{color}
\usepackage{graphicx}
\usepackage[utf8x]{inputenc}
\usepackage[T1]{fontenc}
\usepackage{siunitx}
\usepackage{amstext}
\usepackage{epstopdf}
\usepackage{amsfonts}
\usepackage{amsmath}
\usepackage{multirow}
\usepackage[dvipsnames]{xcolor}

\newcommand\klepiii[1]{{#1}}
\newcommand\klepii[1]{{#1}}
\newcommand\klep[1]{#1}

\begin{document}

%\title{Adaptive negativity estimation with long short-term memory neural networks and collective measurements}

\title{Adaptive Negativity Estimation via Collective Measurements
%processed by Long Short-Term Memory Neural Networks
}

\author{Martin Zeman} \email{martin.zeman01@upol.cz}
\affiliation{Palacký University in Olomouc, Faculty of Science, Joint Laboratory of Optics of Palacký University and Institute of Physics AS CR, 17. listopadu 12, 771 46 Olomouc, Czech Republic}
\affiliation{Institute of Physics of the Academy of Sciences of the Czech Republic, Joint Laboratory of Optics of Palacký University and Institute of Physics AS CR, 17. listopadu 50a, 772 07 Olomouc, Czech Republic}

\author{Vojtěch Tr\'{a}vn\'{i}\v{c}ek}
\email{travnicekv@fzu.cz}
\affiliation{Institute of Physics of the Academy of Sciences of the Czech Republic, Joint Laboratory of Optics of Palacký University and Institute of Physics AS CR, 17. listopadu 50a, 772 07 Olomouc, Czech Republic}

\author{Anton\'{i}n \v{C}ernoch}
\email{acernoch@fzu.cz}
\affiliation{Institute of Physics of the Academy of Sciences of the Czech Republic, Joint Laboratory of Optics of Palacký University and Institute of Physics AS CR, 17. listopadu 50a, 772 07 Olomouc, Czech Republic}

\author{Jan Soubusta} \email{jan.soubusta@upol.cz}
\affiliation{Palacký University in Olomouc, Faculty of Science, Joint Laboratory of Optics of Palacký University and Institute of Physics AS CR, 17. listopadu 12, 771 46 Olomouc, Czech Republic}

\author{Karel Lemr}
\email{k.lemr@upol.cz}
\affiliation{Palacký University in Olomouc, Faculty of Science, Joint Laboratory of Optics of Palacký University and Institute of Physics AS CR, 17. listopadu 12, 771 46 Olomouc, Czech Republic}

\begin{abstract}
This paper explores an efficient method for entanglement quantification in two-qubit and qubit-qutrit quantum systems based upon the framework of collective measurements in conjunction with machine learning. We introduce an adaptive measurement procedure in which measurement settings are dynamically adjusted based on prior measurement outcomes aiming to optimize the inference precision given a limited number of these measurement settings.
The procedure makes use of the Long Short-Term Memory networks to recurrently process collective measurements on two copies of the investigated states. Obtained results demonstrate the tangible benefits of the adaptive measurements in comparison to previously described non-adaptive strategies.
\end{abstract}

\date{\today}
\maketitle

%====================================
\section{Introduction}

Quantum entanglement is a fundamental consequence of the superposition principle applied to multipartite quantum states \cite{Einstein:EPR}. It has been studied for nearly a century \cite{Zeilinger:QT,Pan:swapp,e91}, but despite all efforts, developing a method for its experimentally feasible quantification in general quantum systems remains a challenging and complex problem \cite{Hiesmayr:entanglement,Gharibian:entanglement}. The ultimate aim is to create an approach that minimizes the number of necessary measurements while maximizing the detection accuracy, thus rendering the process more efficient and practical for deployment in quantum technologies.

\begin{figure}[!t!]
\includegraphics[width=.8\linewidth]{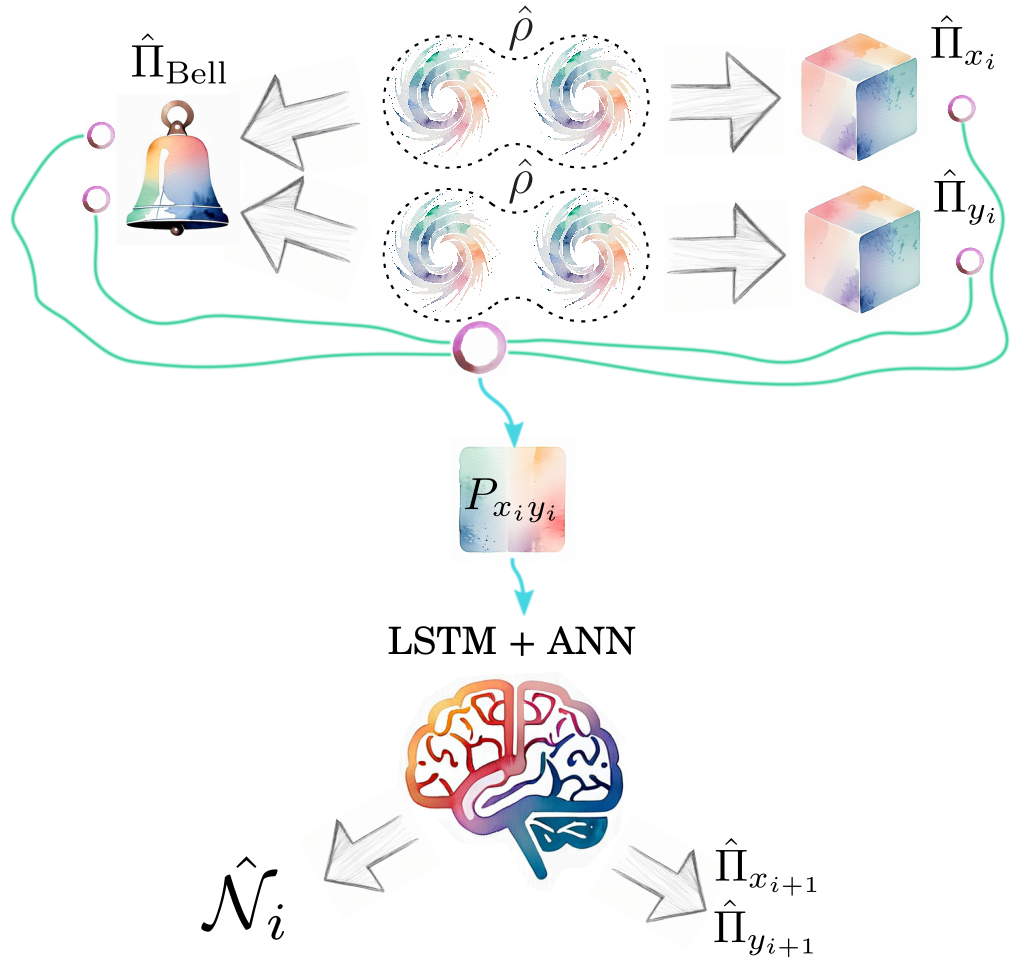}
\caption{\label{fig:coll} The conceptual diagram of a two-copy collective measurement and subsequent negativity prediction by artificial neural network (ANN) model with Long Short-Term Memory (LSTM) cells. Two instances of the system under investigation, denoted as $\hat{\rho}$, are measured simultaneously. During the measurement, one subsystem from each instance undergoes a local projection using operators $\hat{\Pi}_{x_i}$ and $\hat{\Pi}_{y_i}$, while the other two subsystems are non-locally projected using the $\hat{\Pi}_{\mathrm{Bell}}$ operator. The results of these measurements, denoted as $P_{x_{i}y_{i}}$, are sequentially used as inputs for the ANN, which then after each measurement estimates the negativity $\hat{\mathcal{N}}_{i}$ of $\hat{\rho}$ and recommends the most suitable local measurements for the next iteration $\hat{\Pi}_{x_{i+1}}$ and $\hat{\Pi}_{y_{i+1}}$.}
\end{figure}

There is a broad portfolio of entanglement quantifiers among which the quantum state negativity is prominently used for its beneficial characteristics such as being an entanglement monotone and measure \cite{Vidal_2002}.
Negativity extends the Peres-Horodecki criterion, which is a necessary and, in some cases, sufficient condition for the presence of entanglement in quantum systems \cite{Peres:PLA19,Peres:PPT,Jaeger:QIP}. Having access only to single-copy measurements on the tested state, negativity needs to be calculated from the density matrix \cite{horodecki:QE} reconstructed in the process of a complete quantum state tomography
\cite{Jezek:Qtom, Miranowicz:Qtom}. This process is hampered by the unfavorable scaling of measurement configurations that increase quadratically with the Hilbert space dimension.

An alternative approach towards entanglement estimation is based on collective measurements which are simultaneous (and generally nonlocal) measurements implemented on multiple copies of the tested state \cite{PhysRevA.61.062307, PhysRevLett.95.240407, Nature.440, PhysRevLett.107.150502, PhysRevA.86.062329, PhysRevA.94.052334}. It has been shown that by performing a $m$-copy collective measurement, one gets access to the $m$th order moment of the density matrix which is needed to calculate negativity of a $m$-dimensional state \cite{PhysRevLett.89.127902, PhysRevA.65.062320, PhysRevA.66.052315}. The drawback of this procedure is the experimental unattainability of such a complex measurement \cite{PhysRevA.95.022331}. However, one can limit oneself to only two-copy collective measurements and still obtain a reasonably precise negativity estimation while maintaining the number of measurement configurations reduced with respect to the full quantum state tomography \cite{PhysRevApplied.15.054006, ROIK2022128270}. The concept of such collective measurement is depicted in Fig.~\ref{fig:coll}. The approximate estimate for negativity obtained from the two-copy collective measurements can then be obtained by harnessing the universal approximation power of artificial neural networks \cite{HORNIK1989359}.

Moreover, this approach is particularly relevant because the experimental configuration for two-copy collective measurements closely mirrors the entanglement swapping protocol \cite{PhysRevLett.80.3891}. This protocol constitutes a critical mechanism for quantum repeaters and relays \cite{PhysRevLett.81.5932, PhysRevA.66.052307}, which are essential components of quantum communication networks \cite{QCommNet:Book}. Two-copy collective measurements have been used, for example, for quantum relay diagnostics \cite{PhysRevApplied.14.064071}, quantum metrology \cite{Coll.meas} or machine learning \cite{Travnicek:HSD} tasks.

In our previous research, the results of collective measurements were processed by fully connected neural networks to estimate the negativity in two-qubit and qubit-qutrit quantum states \cite{PhysRevApplied.15.054006, ROIK2022128270, PhysRevApplied.15.054006, Travnicek_sensitivity_selectivity,Abo2023,Soubusta2025:130911}. However, the parameters of these measurements were determined in advance and were therefore state independent. In this work, we extend this approach by introducing adaptive measurements, in which the settings of subsequent measurements are dynamically adjusted based on the outcomes of previous measurements.

Adaptive quantum measurements are known to be an efficient tool for extracting useful information from a quantum state \cite{Hentschel}. Their efficiency is measured in term of obtained precision while requiring less resources (fewer measurements, smaller samples or shorter computation time) with respect to non-adaptive approaches. Adaptive methods started being developed for the purposes of the quantum state tomography and estimation \cite{PhysRevA.61.032306, PhysRevA.65.050303, PhysRevA.78.064303}. Originally, the algorithms were based on the Bayes' rule \cite{PhysRevA.85.052120, PhysRevA.93.012103}, which is recently being replaced by neural networks \cite{AdaptiveTomographyNN, PhysRevA.109.012402}. Apart from the quantum state tomography, adaptive measurements outperform their non-adaptive counterparts in quantum phase estimation \cite{PhysRevA.70.043812, linden2025adaptivequantumphaseestimation, PhysRevA.109.042412}, entanglement detection \cite{Lerch:14, Games_preparation} or quantum communications \cite{PhysRevLett.129.120504}.

As we document in later sections, the combination of adaptive and collective measurements leads to a tangible improvement in the precision of negativity estimation for a given number of measurement configurations. The recurrent nature of this task requires a shift from fully connected neural networks to recurrent neural networks (RNN). In this work, we employ the Long Short-Term Memory (LSTM) network, a specialized type of RNN designed to identify and retain patterns in sequential data \cite{LSTM}. LSTM networks are particularly effective in overcoming the vanishing gradient problem \cite{RNN_problems}, which makes them well suited for adaptive quantum measurement tasks. It should be stressed out that once a model is trained, it's daily operation is significantly faster that the algorithms based on the Bayes' rule \cite{AdaptiveTomographyNN}.

%====================================
\section{Methods}
Our goal is to infer the value of negativity from an adaptively iterated series of collective measurements. As indicated in the introduction, we focus on collective measurements applied to two copies of the investigated state. More specifically, to two copies of (i) a two-qubit and (ii) a qubit-qutrit state. This choice is motivated by a prominent role of the negativity which is for these states a necessary and sufficient entanglement witness as well as an entanglement measure \cite{Vidal_2002}.

\begin{figure}[!t!]
\includegraphics{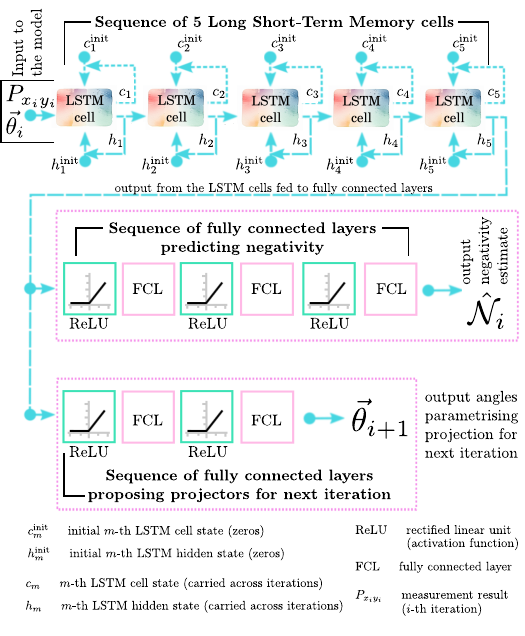}
\caption{\label{fig:network} \klepiii{Topology of the artificial neural network model. In each iteration $i$, the model takes the collective measurement result $P_{x_iy_i}$ and parameters of the corresponding local projectors $\vec{\theta}_i$ as its input. It then processes these values through a sequence of 5 consecutive LSTM cells followed by two sequences of fully connected layers, one rendering the negativity prediction $\hat{\mathcal{N}}_{i}$ and the second proposing the angles  $\vec{\theta}_{i+1}$ for the construction of the local projectors $\hat{\Pi}_{x_{i+1}}$ and $\hat{\Pi}_{y_{i+1}}$ to be applied in the next iteration (see the main text for the parametrization of projectors via $\vec{\theta}$). The LSTM cell states $c_m$ and hidden states $h_m$ ($m = 1,2...5$) are vectors of 128 32-bit float numbers initialized as zeros before the first iteration. Values of $c_m$ and $h_m$ calculated in the $i$-th iteration are stored and applied to the cells in the next iteration constituting thus the necessary memory for the recurrent operation of the LSTM cells. The fully connected layers consist of 256 nodes (each a 32-bit float number). The rectified linear unit function is used as the activation function.}}
\end{figure}

The investigated state is described by its density matrix $\hat{\rho}$. Thus, the two copies of the state can be described simply as a direct product 
\begin{equation}
\label{eq:two_copies}
\hat{\Omega} = \hat{S}^T \hat{\rho} \hat{S} \otimes \hat{\rho},    
\end{equation}
where $\hat{S}$ stands for the operator swapping subsystems in the density matrix of the first copy ($\hat{\rho}_{AB} \xrightarrow{\hat{S}} \hat{\rho}_{BA}$). The role of the swapping operator $\hat{S}$ will become evident in subsequent derivations.

The collective measurement itself consists of a local and nonlocal part. \klepiii{The local part is described by two local projectors $\hat{\Pi}_{x}$ and $\hat{\Pi}_{y}$ acting on the same subsystems across the two copies (see conceptual diagram in Fig.~\ref{fig:coll}). These are pairs of projection operators onto any pure qubit or qutrit state and are parametrized by pairs of two \klepii{($\cos\theta_1 |0\rangle + \mathrm{e}^{i\theta_2} \sin\theta_1 |1\rangle$)} and four angles \klepii{($\cos\theta_1 |0\rangle + \mathrm{e}^{i\theta_3} \sin\theta_1 \cos\theta_2 |1\rangle + \mathrm{e}^{i\theta_4} \sin\theta_1 \sin\theta_2 |2\rangle$)} for qubits and qutrits, respectively; hereinafter referred to as $\vec{\theta}$. The angle parametrization proves particularly convenient, as any set of angles returned by the neural network model corresponds to a physically valid projector.} The remaining two subsystems are simultaneously projected onto a common entangled (non-local) state denoted by the projector $\hat{\Pi}_{\mathrm{Bell}}$. We chose the projector on the fully anti-symmetric two-qubit subspace
\begin{equation}
    \hat{\Pi}_{\mathrm{Bell}} = |\Psi^-\rangle\langle\Psi^-|,\quad |\Psi^-\rangle = \frac{1}{\sqrt{2}} \left(|01\rangle - |10\rangle\right)
\end{equation}
to be the fixed non-local projector. In case of the qubit-qutrit states, the non-local projection is performed to the qubit subsystems. The outcome of the collective measurement is an overall projection probability expressed as
\begin{equation}
\label{eq:collective_probability}
   P_{xy} = \mathrm{Tr}[\hat{\Omega}
   (\hat{\Pi}_{x}\otimes\hat{\Pi}_{\mathrm{Bell}}
                   \otimes\hat{\Pi}_{y})].
\end{equation}
This equation illustrates how the swap operator $\hat{S}$ ensures correct ordering of the subsystems. Typically, one chooses a single nonlocal projector $\hat{\Pi}_{\mathrm{Bell}}$ (or POVM operator in a more general case) and repeats the measurement for a sequence of different local projectors $\lbrace\hat{\Pi}_{x_i} \otimes \hat{\Pi}_{y_i}\rbrace$. Thus, one obtains a sequence of probabilities $\lbrace P_{x_iy_i}\rbrace$. Negativity or other quantity of interest is then inferred from this sequence. Note that the probabilities $P_{xy}$ are experimentally accessible since Eq. (\ref{eq:collective_probability}) can be directly implemented.

The sequence of local operators is fixed in the case of a non-adaptive procedure. In this paper, we present an adaptive procedure, in which the local projectors imposed on the state in the $(i+1)$th iteration step are proposed based on the results of the $i$ previous measurements. We make use of a neural network model consisting of both recurrent LSTM cells and fully connected layers to estimate negativity from the the total of $i$ collective measurements ($P_{x_1y_1}, P_{x_2y_2}, ..., P_{x_iy_i}$) and also to predict the optimal local projectors for the $(i+1)$th iteration ($\hat{\Pi}_{x_{i+1}}$ and $\hat{\Pi}_{y_{i+1}}$). The entire neural network model is schematically visualized in Fig.~\ref{fig:network}. \klepiii{In each iteration $i$, the model processes the outcome of the collective measurement $P_{x_iy_i}$ along with parameters $\vec{\theta}_i$ of the projectors used to acquire it. First, the input is subjected to a sequence of five consecutive LSTM cells. Their cell states and hidden states are initialized as vectors of zeros before the first iteration and subsequently their values calculated in the $i$-th iteration are carried over to the next one, thus representing the memory needed for the recurrent operation of the network. The output of the last LSTM cell is used as input to two sequences of fully connected layers: one to estimate negativity and the second to propose projectors to be applied in the next iteration. Note that the entire model is run in each iteration.}

We label $\hat{\mathcal{N}}_i(\hat{\rho})$ the \emph{negativity estimate} of state $\hat{\rho}$ after the total of $i$ collective measurements have been performed. For purposes of training and testing of our models, we analytically calculate the \emph{true negativity}
\begin{equation}
   \mathcal{N}(\hat{\rho}) = \Bigg |\sum_{\lambda_{j}<0}\lambda_{j}\Bigg |,
\end{equation}
where $\lambda_{j}$ denote the eigenvalues of a partially transposed density matrix $\hat{\rho}^\Gamma$.

%====================================
\subsection{Training and testing of models}
The training and testing of the models started by generating random two-qubit or qubit-qutrit states using the method described in \cite{Maziero_2015}. Each model is trained on a set of $2^{18}$ random states that were split into mini-batches $\mathcal{B}$ of size $B$. The input features of the model consist of the collective measurement probabilities $P_{xy}$ obtained by applying  Eq. (\ref{eq:collective_probability}) to each density matrix $\hat{\Omega}$. One input feature per density matrix is calculated in each iteration of the model (labeled $i = 1, ..., n$\klepii{, with $n$ denoting the total number of iterations}). In the first iteration, a fixed set of local projectors $\Pi_{x_1} = \Pi_{y_1} = |0\rangle\langle 0|$ is applied. For the next iterations $i = 2, ..., n$, the model proposes new pairs of local projectors ($\Pi_{x_i}$ and $\Pi_{y_i}$) independently for every density matrix. \klepii{The neural network outputs a vector of float values representing the angles $\vec{\theta}$, which are then used to construct the local projectors for the next iteration via the definitions above.} In addition to the newly proposed local projectors, in every iteration, the model outputs the negativity estimate $\hat{\mathcal{N}}_i(\hat{\rho})$ corresponding to each investigated state. \klepii{The negativity estimate is directly output by the network as a single float. The reader is encouraged to inspect the source code available on Figshare \citep{figshare} for further implementation details.}

Once all iterations $i = 2, ..., n$ are complete for one mini-batch of density matrices, back-propagation and parameter updates are performed using the Adam optimizer with a learning rate of $0.001$. We implemented two different loss functions, serving as figures of merit during training. Based on the selected loss function, we distinguish the \emph{greedy} approach
\begin{equation}
    \label{eq:loss_greedy}
    \mathcal{L}^{(2)}_\mathrm{greedy}(n) = \frac{1}{nB}\sum_{\hat{\rho}\in\mathcal{B}} \sum_{i=1}^{n} \left(\hat{\mathcal{N}}_{i}(\hat{\rho}) - \mathcal{N}(\hat{\rho}) \right)^2
\end{equation}
and the \emph{last} approach
\begin{equation}
    \label{eq:loss_last}
    \mathcal{L}^{(2)}_\mathrm{last}(n) = \frac{1}{B}\sum_{\hat{\rho}\in\mathcal{B}} \left(\hat{\mathcal{N}}_{n}(\hat{\rho}) - \mathcal{N}(\hat{\rho}) \right)^2.
\end{equation}
Both loss functions are ordinary $\mathcal{L}^{(2)}$ distances between the predicted and true negativity values. The difference is that, in the case of the \emph{greedy} algorithm, the model is optimized based on its intermediate negativity predictions made in each iteration. Conversely, models trained using the \emph{last} approach are only optimized based on the final negativity prediction. The \emph{greedy} models' parameters generally converge faster, and the models are easier to train. \emph{Last} models are more difficult to train and fail more frequently during training. However, if successful, they offer better final negativity estimates. This procedure is repeated for every mini-batch, thus completing the training epoch.

To prevent overfitting, a separate validation set of $2^{16}$ random density matrices is generated and used to assess the model performance after each epoch. If no improvement on the validation set is detected for 10 consecutive epochs, the size of the mini-batches increases by a factor of two starting from $B=32$ until it reaches its maximum value $B=512$ \cite{smith2018dontdecaylearningrate}. Training is then terminated and the best-performing model is saved for testing. The models are tested on a third and independent set of $2^{16}$ random density matrices. The results are presented in subsequent sections.

The entire procedure described was coded in the Python programming language using the PyTorch library \cite{paszke2019pytorchimperativestylehighperformance} to implement the models and their training. All programming code, final trained models and testing results can be accessed on Figshare \cite{figshare}.

%====================================
\section{Results}
As indicated in the Methods section, two different approaches based on two different loss functions designated \emph{greedy} (\ref{eq:loss_greedy}) and \emph{last} (\ref{eq:loss_last}) were tested. On top of that, we trained the neural networks using an optimal fixed (\emph{non-adaptive}) set of bases. Non-adaptive training and operation of the models was simply achieved by overriding the usage of model-proposed basis in every iteration by a fixed basis from a list. These bases and their ordering were chosen according to previous research \cite{Travnicek_sensitivity_selectivity,Soubusta2025:130911}. Comparing the performance of adaptive models with models trained with a non-adaptive set of bases allows quantifying the improvement gained by the models' adaptability.

We have selected two quantifiers to assess the performance of the models. Straightforwardly, one can make use of the $\mathcal{L}^{(1)}$ distance between the true and estimated negativity averaged over the entire testing set $\mathcal{T}$ (of size $T$)
\begin{equation}
    \label{eq:l1}
    \mathcal{L}^{(1)}(n) = \frac{1}{T}\sum_{\hat{\rho}\in\mathcal{T}} |\hat{\mathcal{N}}_{n}(\hat{\rho}) - \mathcal{N}(\hat{\rho}) |.
\end{equation}
 Unlike the $\mathcal{L}^{(2)}$ distance used for training, the $\mathcal{L}^{(1)}$ distance has a simpler physical interpretation as the deviation of the predicted negativity from its true value. The second quantifier used is the coefficient of determination $R^2$ describing the variance of the estimated values of negativity with respect to the true values
 \begin{equation}
     \label{eq:r2}
     R^2(n) = 1-\frac{\sum_{\hat{\rho}\in\mathcal{T}} \left(\hat{\mathcal{N}}_{n}(\hat{\rho}) - \mathcal{N}(\hat{\rho}) \right)^2}{\sum_{\hat{\rho}\in\mathcal{T}} \left(\hat{\mathcal{N}}_{n}(\hat{\rho}) - \langle \hat{\mathcal{N}}_{n}\rangle \right)^2},
 \end{equation}
where $\langle \hat{\mathcal{N}}_{n}\rangle$ is the predicted value of negativity averaged over the training set.

For every set of parameters such as the number of iterations or the loss function type, a series of at least five models were trained to obtain the performance statistics. A model is deemed successfully trained if, for a given number of iterations $n$, it performs better in terms of the $\mathcal{L}^{(1)}$ distance then any other model trained on less iterations. Otherwise the model is discarded. We present both the mean performance over the entire set of successful models as well as the performance of the best model \klep{(labelled as ``mean'' and ``best'' in the subsequent tables)}.

%====================================
\subsection{Two-qubit states}

First, we present the results concerning two-qubit states. The qubit-qutrit systems are discussed afterward. Tables \ref{tab:QB_AG} and \ref{tab:QB_AL} summarize the performance statistics of the models trained using the \emph{greedy} and \emph{last} approaches, respectively. The rows correspond to the number of iterations $n$ ranging from 2 to 10; the case $n=1$ is excluded due to its incompatibility with the adaptive measurement framework. The upper limit of $n = 10$ measurement configurations corresponds to the maximum number of independent collective measurements available for a two-qubit system \cite{Soubusta2025:130911}, which follows directly from the symmetry of the density matrix $\hat{\Omega}$.

\begin{table}
  \caption{\label{tab:QB_AG} 
  Numerical results for two-qubit systems using an adaptive \emph{greedy} measurement strategy as a function of the number of $n$ iterations. The figures of merit $\mathcal{L}^{(1)}$ and $R^2$ are defined in the main text.}
\begin{ruledtabular} 
 \begin{tabular}{ccccc}
  $n$  & {$\mathcal{L}^{(1)}\ \text{mean}$} & $\mathcal{L}^{(1)}\ \text{best}$ & {$\text{$R^2$ mean}$} & {$\text{$R^2$ best}$} \\
  \hline
   2 & 0.0900(03) & 0.0895 & 0.5952(67) & 0.6003 \\
   3 & 0.0405(02) & 0.0402 & 0.9070(13) & 0.9095 \\
   4 & 0.0315(11) & 0.0302 & 0.9419(67) & 0.9470 \\
   5 & 0.0225(04) & 0.0217 & 0.9690(11) & 0.9700 \\
   6 & 0.0179(03) & 0.0173 & 0.9806(09) & 0.9818 \\
   7 & 0.0146(16) & 0.0125 & 0.9869(25) & 0.9900 \\
   8 & 0.0118(02) & 0.0114 & 0.9912(04) & 0.9915 \\
   9 & 0.0105(06) & 0.0097 & 0.9929(07) & 0.9940 \\
  10 & 0.0092(04) & 0.0083 & 0.9942(05) & 0.9953
 \end{tabular}
\end{ruledtabular}
\end{table}

\begin{table}
  \caption{\label{tab:QB_AL} 
  Numerical results for two-qubit systems using an adaptive \emph{last} measurement strategy as a function of the number of $n$ iterations. The figures of merit $\mathcal{L}^{(1)}$ and $R^2$ are defined in the main text.}
\begin{ruledtabular} 
 \begin{tabular}{ccccc}
  $n$  & {$\mathcal{L}^{(1)}\ \text{mean}$} & $\mathcal{L}^{(1)}\ \text{best}$ & {$\text{$R^2$ mean}$} & {$\text{$R^2$ best}$} \\
  \hline
   2 & 0.0902(04) & 0.0896 & 0.5969(47) & 0.5896 \\
   3 & 0.0401(02) & 0.0399 & 0.9077(16) & 0.9084 \\
   4 & 0.0248(18) & 0.0238 & 0.9628(41) & 0.9651 \\
   5 & 0.0180(29) & 0.0145 & 0.9801(59) & 0.9871 \\
   6 & 0.0111(05) & 0.0105 & 0.9920(07) & 0.9929 \\
   7 & 0.0096(03) & 0.0092 & 0.9937(05) & 0.9942 \\
   8 & 0.0083(01) & 0.0082 & 0.9951(02) & 0.9952 \\
   9 & 0.0076(02) & 0.0074 & 0.9955(02) & 0.9958 \\
  10 & 0.0070(01) & 0.0068 & 0.9960(01) & 0.9961
 \end{tabular}
\end{ruledtabular}
\end{table}

\begin{figure}[ht]
  \includegraphics[width=.8\linewidth]{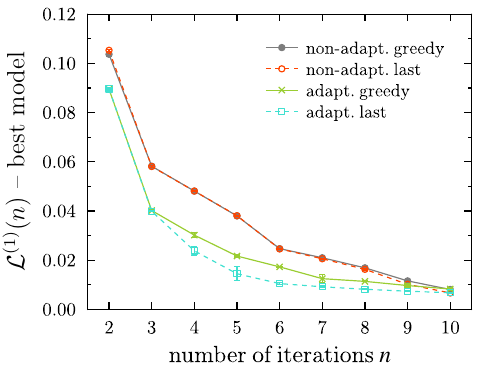}
  \caption{\label{fig:QB_results} Comparison of the ANN models for two-qubit systems. Performance in terms of the $\mathcal{L}^{(1)}$ error for all combinations of adaptive/non-adaptive and greedy/last approaches is visualized as a function of the number of $n$ iterations.}
\end{figure}

\begin{figure*}[bt]
  \includegraphics[width=\linewidth]{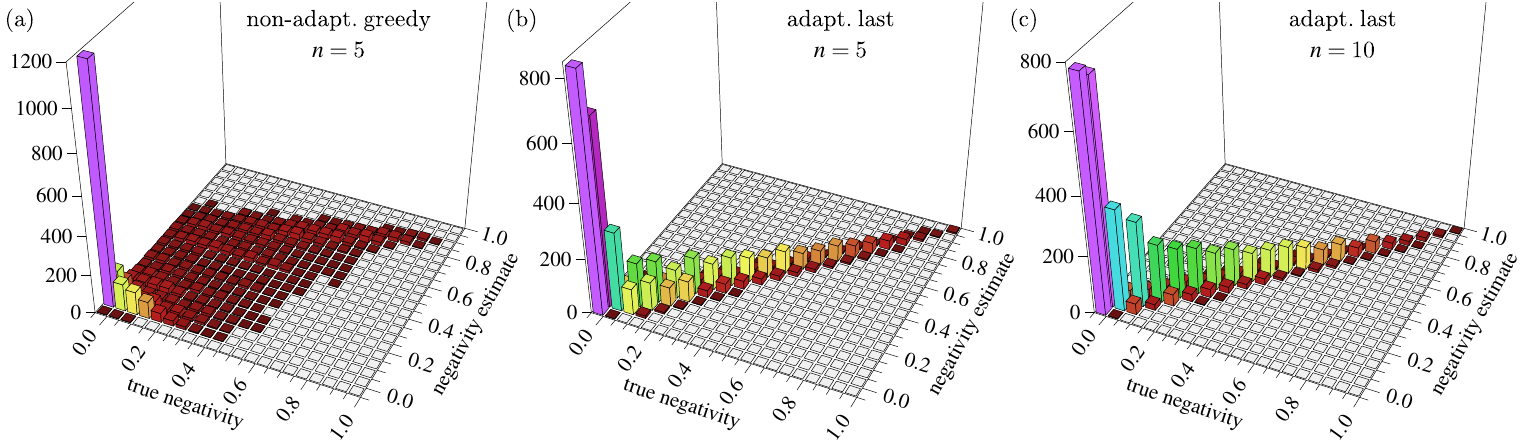}
  \caption{\label{fig:QB_histograms} Histograms showing the distribution of paired values of the true negativity $\mathcal{N}$ and its estimate by the models $\hat{\mathcal{N}}$ in two-qubit systems: (a) the non-adaptive strategy with $n=5$ iterations, (b) the adaptive $\emph{last}$ approach with $n=5$ iterations and (c) the adaptive $\emph{last}$ approach with $n=10$ iterations. Note that ideally only diagonal elements, where $\mathcal{N} = \hat{\mathcal{N}}$, should have non-zero values.}
\end{figure*}

For comparison, the performance of equivalent non-adaptive models using an optimal sequence of fixed bases is summarized in Tables \ref{tab:QB_FG} and \ref{tab:QB_FL}. The performance of the best adaptive and non-adaptive models, as described by $\mathcal{L}^{(1)}(n)$, is visualized in Fig.~\ref{fig:QB_results} as a function of the number of iterations $n$. Additionally, we plot two-dimensional histograms of the estimated and true negativity values in Fig.~\ref{fig:QB_histograms} for several selected parameters. These histograms were obtained by processing samples of 4096 random states. The separable states (zero valued negativity) are accumulated in the first bin.

\begin{table}
  \caption{\label{tab:QB_FG} 
  Numerical results for two-qubit systems using a non-adaptive optimal basis set and the \emph{greedy} estimation approach. Here, $n$ denotes the number of iterations. The figures of merit $\mathcal{L}^{(1)}$ and $R^2$ are defined in the main text.}
\begin{ruledtabular} 
 \begin{tabular}{ccccc}
  $n$  & {$\mathcal{L}^{(1)}\ \text{mean}$} & $\mathcal{L}^{(1)}\ \text{best}$ & {$\text{$R^2$ mean}$} & {$\text{$R^2$ best}$} \\
  \hline
   2 & 0.1049(08) & 0.1037 & 0.5363(58) & 0.5406 \\
   3 & 0.0588(05) & 0.0582 & 0.8124(21) & 0.8112 \\
   4 & 0.0484(02) & 0.0481 & 0.8702(14) & 0.8720 \\
   5 & 0.0382(03) & 0.0380 & 0.9152(03) & 0.9155 \\
   6 & 0.0250(02) & 0.0247 & 0.9616(06) & 0.9623 \\
   7 & 0.0211(00) & 0.0210 & 0.9721(07) & 0.9728 \\
   8 & 0.0175(04) & 0.0169 & 0.9800(05) & 0.9808 \\
   9 & 0.0119(03) & 0.0116 & 0.9909(06) & 0.9916 \\
  10 & 0.0084(03) & 0.0081 & 0.9951(03) & 0.9953
 \end{tabular}
\end{ruledtabular}
\end{table}

\begin{table}
  \caption{\label{tab:QB_FL} 
  Numerical results for two-qubit systems using a non-adaptive optimal basis set and the \emph{last} estimation approach. Here, $n$ denotes the number of iterations. The figures of merit $\mathcal{L}^{(1)}$ and $R^2$ are defined in the main text.}
\begin{ruledtabular} 
 \begin{tabular}{ccccc}
  $n$  & {$\mathcal{L}^{(1)}\ \text{mean}$} & $\mathcal{L}^{(1)}\ \text{best}$ & {$\text{$R^2$ mean}$} & {$\text{$R^2$ best}$} \\
  \hline
   2 & 0.1059(05) & 0.1053 & 0.5331(65) & 0.5257 \\
   3 & 0.0588(04) & 0.0581 & 0.8144(16) & 0.8161 \\
   4 & 0.0483(02) & 0.0481 & 0.8712(23) & 0.8709 \\
   5 & 0.0382(01) & 0.0381 & 0.9155(10) & 0.9153 \\
   6 & 0.0249(02) & 0.0246 & 0.9619(05) & 0.9625 \\
   7 & 0.0208(02) & 0.0206 & 0.9725(03) & 0.9729 \\
   8 & 0.0166(03) & 0.0163 & 0.9814(04) & 0.9821 \\
   9 & 0.0105(02) & 0.0102 & 0.9923(04) & 0.9928 \\
  10 & 0.0068(01) & 0.0067 & 0.9960(02) & 0.9961
 \end{tabular}
\end{ruledtabular}
\end{table}

Our results indicate that, as expected, the adaptive models outperform their non-adaptive counterparts across the entire range of iteration steps until
$n=10$. The convergence at $n=10$ is expected since this corresponds to the maximum number of unbiased measurement configurations. Moreover, the adaptive models trained using the \emph{last} approach outperform the \emph{greedy} models as they minimize the error solely at the last iteration step. In contrast, the \emph{greedy} models are optimized to provide reasonable estimates at every iteration. Conversely, we did not observe any significant difference in performance between the \emph{greedy} and \emph{last} non-adaptive models.

The presented results highlight the tangible benefits of using the adaptive measurement strategy. For example, at $n=5$ the average deviation between the true negativity and its estimate made by the best available adaptive model is only $0.0145$, which is more then $2.5\times$ smaller than that of the best available non-adaptive model ($0.0380$). It should also be noted that negativity cannot be analytically calculated from two-copy collective measurements. Observing estimates deviating by less then $1.5~\%$ while using only half of the total number of unbiased measurements demonstrates the effectiveness of the ANN-based models.

\subsection{Qubit-qutrit states}

Similarly to the two-qubit states, we trained and tested models for negativity estimation also on qubit-qutrit systems. In this case, the qubit subsystems of the two copies undergo the non-local $\hat\Pi_\mathrm{Bell}$ projection and the qutrit subsystems are locally projected. This choice is warranted by the uniqueness of the singlet two-qubit Bell state (the sole anti-symmetric fully entangled state). In this geometry, there are up to 45 unbiased local projections \cite{Travnicek_sensitivity_selectivity}. However, we limit our investigation to $n=21$ since for a larger number of iterations the training becomes too complex at least given our computation power. One can also argue that there is no point in using two-copy collective measurements if the number of measurements exceeds half of the measurements necessary for standard quantum state tomography. Quantum state tomography of a qubit-qutrit system can be accomplished with 36 measurements.

\begin{figure}[th]
  \includegraphics[width=.8\linewidth]{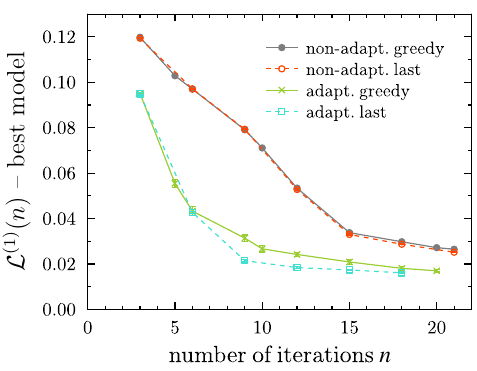}
  \caption{\label{fig:QT_results} Comparison of the ANN models for qubit-qutrit systems. Performance in terms of the $\mathcal{L}^{(1)}$ error for all combinations of adaptive/non-adaptive and greedy/last approaches is visualized as a function of the number of $n$ iterations.}
\end{figure}

\begin{figure*}[bt]
  \includegraphics[width=\linewidth]{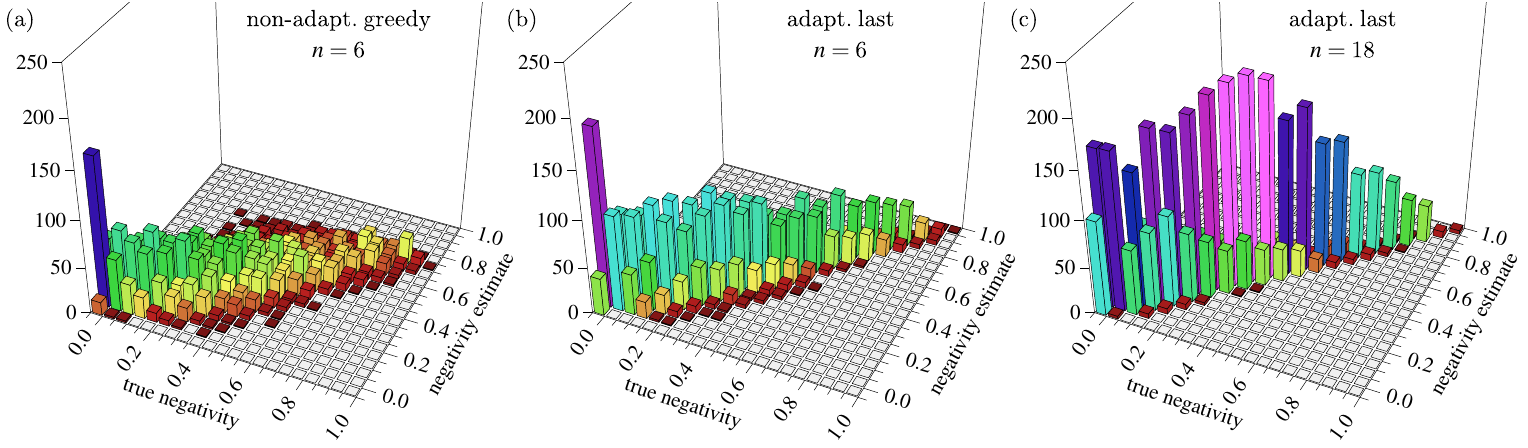}
  \caption{\label{fig:QT_histograms} Histograms showing the distribution of paired values of the true negativity $\mathcal{N}$ and its estimate by the models $\hat{\mathcal{N}}$ in qubit-qutrit systems: (a) the non-adaptive strategy with $n=6$ iterations, (b) the adaptive $\emph{last}$ approach with $n=6$ iterations and (c) the adaptive $\emph{last}$ approach with $n=18$ iterations. Note that ideally only diagonal elements, where $\mathcal{N} = \hat{\mathcal{N}}$, should have non-zero values.}
\end{figure*}

Using the same two quantifiers $\mathcal{L}^{(1)}$ and $R^2$, we assess the performance of the models on the qubit-qutrit states. Numerical values are presented in Tables \ref{tab:QT_AG} and \ref{tab:QT_AL} for the \emph{greedy} and \emph{last} approaches respectively. Tables \ref{tab:QT_FG} and \ref{tab:QT_FL} provide performances of their non-adaptive counterparts. The resulting trend in terms of the $\mathcal{L}^{(1)}(n)$ measure as a function of the number of iterations $n$ is depicted in Fig.~\ref{fig:QT_results}. Two-dimensional histograms for a selection of parameters are shown in Fig.~\ref{fig:QT_histograms}. Histograms were generated using the same procedure as described in the preceding subsection dedicated to the two-qubit states.

The observations on qubit-qutrit systems are very similar to those of two-qubit systems. The adaptive strategies clearly outperform the non-adaptive approaches. This can be numerically documented, for instance, by considering the best-performing adaptive model for $n=12$ iterations, which makes estimates that on average deviate from true negativity values by $0.0184$. The non-adaptive model makes errors almost three times bigger ($0.0528$). Adaptive models trained using the \emph{last} approach again outperform the \emph{greedy} models. No significant difference exists among their non-adaptive equivalents.

\begin{table}
  \caption{\label{tab:QT_AG} 
  Numerical results for qubit-qutrit systems using an adaptive \emph{greedy} measurement strategy as a function of the number of $n$ iterations. The figures of merit $\mathcal{L}^{(1)}$ and $R^2$ are defined in the main text.
  }
\begin{ruledtabular} 
 \begin{tabular}{ccccc}
  $n$  & {$\mathcal{L}^{(1)}\ \text{mean}$} & $\mathcal{L}^{(1)}\ \text{best}$ & {$\text{$R^2$ mean}$} & {$\text{$R^2$ best}$} \\
  \hline
   3 & 0.0953(02) & 0.0951 & 0.7059(33) & 0.7084 \\
   5 & 0.0581(18) & 0.0555 & 0.8835(49) & 0.8907 \\
   6 & 0.0451(21) & 0.0433 & 0.9269(63) & 0.9323 \\
   9 & 0.0329(14) & 0.0314 & 0.9603(34) & 0.9641 \\
  10 & 0.0284(14) & 0.0267 & 0.9702(31) & 0.9732 \\
  12 & 0.0251(08) & 0.0242 & 0.9767(12) & 0.9780 \\
  15 & 0.0218(10) & 0.0209 & 0.9819(16) & 0.9836 \\
  18 & 0.0194(09) & 0.0181 & 0.9857(12) & 0.9876 \\
  20 & 0.0178(05) & 0.0170 & 0.9878(06) & 0.9888
 \end{tabular}
\end{ruledtabular}
\end{table}

\begin{table}
  \caption{\label{tab:QT_AL} 
  Numerical results for qubit-qutrit systems using an adaptive \emph{last} measurement strategy as a function of the number of $n$ iterations. The figures of merit $\mathcal{L}^{(1)}$ and $R^2$ are defined in the main text.
  }
\begin{ruledtabular} 
 \begin{tabular}{ccccc}
  $n$  & {$\mathcal{L}^{(1)}\ \text{mean}$} & $\mathcal{L}^{(1)}\ \text{best}$ & {$\text{$R^2$ mean}$} & {$\text{$R^2$ best}$} \\
  \hline
   3 & 0.0954(05) & 0.0949 & 0.7054(16) & 0.7039 \\
   6 & 0.0435(05) & 0.0427 & 0.9311(22) & 0.9354 \\
   9 & 0.0220(06) & 0.0215 & 0.9812(11) & 0.9820 \\
  12 & 0.0192(05) & 0.0184 & 0.9856(07) & 0.9870 \\
  15 & 0.0176(02) & 0.0174 & 0.9878(03) & 0.9878 \\
  18 & 0.0168(04) & 0.0161 & 0.9890(05) & 0.9895
 \end{tabular}
\end{ruledtabular}
\end{table}

\begin{table}
  \caption{\label{tab:QT_FG} 
  Numerical results for qubit-qutrit systems using a non-adaptive optimal basis set and the \emph{greedy} estimation approach. Here, $n$ denotes the number of iterations. The figures of merit $\mathcal{L}^{(1)}$ and $R^2$ are defined in the main text.
  }
\begin{ruledtabular} 
 \begin{tabular}{ccccc}
  $n$  & {$\mathcal{L}^{(1)}\ \text{mean}$} & $\mathcal{L}^{(1)}\ \text{best}$ & {$\text{$R^2$ mean}$} & {$\text{$R^2$ best}$} \\
  \hline
   3 & 0.1200(02) & 0.1198 & 0.5665(25) & 0.5698 \\
   5 & 0.1034(03) & 0.1029 & 0.6660(15) & 0.6678 \\
   6 & 0.0973(02) & 0.0970 & 0.6993(14) & 0.7002 \\
   9 & 0.0796(03) & 0.0791 & 0.7798(15) & 0.7809 \\
  10 & 0.0712(02) & 0.0711 & 0.8246(10) & 0.8247 \\
  12 & 0.0537(02) & 0.0534 & 0.9011(04) & 0.9012 \\
  15 & 0.0342(03) & 0.0338 & 0.9580(05) & 0.9583 \\
  18 & 0.0300(01) & 0.0298 & 0.9675(05) & 0.9679 \\
  20 & 0.0274(02) & 0.0272 & 0.9726(04) & 0.9731 \\
  21 & 0.0268(02) & 0.0265 & 0.9737(03) & 0.9740
 \end{tabular}
\end{ruledtabular}
\end{table}

\begin{table}
  \caption{\label{tab:QT_FL} 
  Numerical results for qubit-qutrit systems using a non-adaptive optimal basis set and the \emph{last} estimation approach. Here, $n$ denotes the number of iterations. The figures of merit $\mathcal{L}^{(1)}$ and $R^2$ are defined in the main text.
  }
\begin{ruledtabular} 
 \begin{tabular}{ccccc}
  $n$  & {$\mathcal{L}^{(1)}\ \text{mean}$} & $\mathcal{L}^{(1)}\ \text{best}$ & {$\text{$R^2$ mean}$} & {$\text{$R^2$ best}$} \\
  \hline
   3 & 0.1196(02) & 0.1194 & 0.5669(63) & 0.5739 \\
   6 & 0.0975(02) & 0.0971 & 0.6984(20) & 0.6970 \\
   9 & 0.0795(02) & 0.0793 & 0.7795(17) & 0.7798 \\
  12 & 0.0531(03) & 0.0528 & 0.9027(11) & 0.9044 \\
  15 & 0.0332(02) & 0.0330 & 0.9598(04) & 0.9601 \\
  18 & 0.0288(01) & 0.0287 & 0.9697(03) & 0.9700 \\
  21 & 0.0254(02) & 0.0252 & 0.9761(04) & 0.9762
 \end{tabular}
\end{ruledtabular}
\end{table}

\klepii{\subsection{State-dependent performance}}

\klep{
The histograms in Figs. \ref{fig:QB_histograms} and \ref{fig:QT_histograms} suggest that the performance of our method is, to some extent, state-dependent. There are indeed states for which the negativity is predicted more accurately than for other states. We have investigated this aspect and quantified it in Tab. \ref{tab:percentiles}. This table presents the percentile values of the $\mathcal{L}^{(1)}$ error for two selected  qubit-qubit models and two qubit-qutrit models. Both models implement the \emph{last} adaptive strategy and are the best performing models found for the respective numer of iterations $n$. A percentile value $\mathcal{L}^{(1)}_{P}$ marks the maximum difference between the predicted and true negativity that can be achieved for $P$ percent of the tested states. $\mathcal{L}^{(1)}_{100}$ is therefore the worst-case scenario. We found that the method typically struggles with partially entangled states ($\mathcal{N}<0.3$) and becomes more precise when predicting the negativity of highly entangled states ($\mathcal{N}>0.8$). This trend is visible in the last row of Tab. \ref{tab:percentiles}, which corresponds to the worst-case scenario among states with negativity $\mathcal{N}>0.9$, and is also consistent with the results on non-adaptive strategies published in Ref. \cite{ROIK2022128270}.

\klepii{Although negativity cannot be calculated analytically from two-copy collective measurements \cite{PhysRevA.95.022331}, other quantum correlation quantifiers—such as Bell nonlocality, entropic witnesses, and the fully entangled fraction (FEF)—can be \cite{PhysRevA.95.030102}. Numerical analysis in Ref. \cite{PhysRevA.95.030102} suggests that the correlation between these measures and negativity increases alongside the amount of entanglement. Specifically, highly entangled states demonstrate a robust correlation between negativity and analytically accessible metrics. Consequently, predicting negativity is significantly easier for highly entangled states, as its value maps much more reliably to the outcomes of two-copy collective measurements. This effect can easily be demonstrated on the case of an almost maximally entangled two-qubit state $\hat{\rho} = (1-\epsilon)|\Phi\rangle\langle\Phi| + \epsilon\frac{1}{4}$, where $|\Phi\rangle$ is an arbitrary maximally entangled two-qubit state and $\epsilon\ll 1$. Note the structural similarity between two-copy collective measurements and the entanglement swapping protocol as it can be identified from the conceptual diagram in Fig.~\ref{fig:coll} as well as from the non-local projection used in Eq. (\ref{eq:collective_probability}). Here, two copies of a tested state $\hat{\rho}_{AB}\otimes \hat{\rho}_{A'B'}$ (where $A, A', B, B'$ denote the respective qubits) are projected via the non-local projector $|\Psi^-\rangle\langle\Psi^-|$ acting on $B,B'$
\begin{equation}
\label{eq:highly}
\hat{\rho}_{AB}\otimes \hat{\rho}_{A'B'} \xrightarrow{|\Psi^-\rangle\langle\Psi^-|_{BB'}} \hat{\rho}'_{AA'} \approx |\Psi^-\rangle\langle\Psi^-|
\end{equation}
The output state from this non-local projection is always close to the singlet Bell state $|\Psi^-\rangle$, regardless of the local properties of the original maximally entangled state $|\Phi\rangle$. The FEF of $\hat{\rho}$, for instance, can then be by approximately obtained in terms of the overlap between the projected state and singlet state
\begin{equation}
1- \mathrm{FEF} \approx \frac{1}{2}(1-\langle\Psi^-|\hat{\rho}'_{AA'}|\Psi^-\rangle).
\end{equation}
This overlap can indeed be measured using a suitable set of local projectors $\hat{\Pi}_x$ and $\hat{\Pi}_y$, since the operator $|\Psi^-\rangle\langle\Psi^-|$ admits a local decomposition \cite{sanctuary2009separationbellstates} that the adaptive neural network is capable of identifying. Consequently the capacity to predict the FEF from two-copy collective measurements and the strong correlation between negativity and the FEF allows the neural network to predict the negativity very precisely in highly entnagled states. Furthermore, note that for partially entangled states, the approximation in Eq. (\ref{eq:highly}) no longer holds, and one cannot simply assume the state $\hat{\rho}_{AA'}$ to be close to a singlet Bell state. At this point, the adaptive strategies outperform their non-adaptive counterparts as evident from the results (see Figs. \ref{fig:QB_results} and \ref{fig:QT_results}).}

\begin{table}
  \caption{\label{tab:percentiles}
  \klep{
  Percentile values of the $\mathcal{L}^{(1)}$ errors, quantifying the state-dependent aspect of two selected models for qubit-qubit and qubit-qtrit states. A percentile value $\mathcal{L}^{(1)}_{P}$ marks the maximum difference between the predicted and true negativity that can be achieved for $P$ percent of the tested states. The last row corresponds to the worst-case scenario limited only to states of negativity $\mathcal{N}>0.9$.}
  }
\begin{ruledtabular} 
 \begin{tabular}{ccccc}
    & \multicolumn{2}{c}{qubit-qubit} &  \multicolumn{2}{c}{qubit-qutrit}\\
   $n$ & 5 & 10 & 6 & 18\\
  \hline
   $\mathcal{L}^{(1)}_{50}$ & 0.0075 & 0.0025 & 0.0324 & 0.0107 \\
   $\mathcal{L}^{(1)}_{90}$ & 0.0397 & 0.0190 & 0.0958 & 0.0387 \\
   $\mathcal{L}^{(1)}_{95}$ & 0.0525 & 0.0272 & 0.1194 & 0.0496 \\
   $\mathcal{L}^{(1)}_{99}$ & 0.0827 & 0.0514 & 0.1721 & 0.0727 \\
   $\mathcal{L}^{(1)}_{100}$ & 0.2172 & 0.1915 & 0.3818 & 0.1724 \\ \hline
   $\mathcal{L}^{(1)}_{100}$ ($\mathcal{N}>0.9$) & 0.0160 & 0.0076 & 0.0339 & 0.0158 \\
 \end{tabular}
\end{ruledtabular}
\end{table}
}

\klepii{
\subsection{Robustness against detection noise}
In experimental practice, the measurements described by Eq.~(\ref{eq:collective_probability}) must be carried out using a large number of instances of $\hat{\Omega}$ to estimate the probabilities with sufficient precision. A finite number of state copies $\hat{\Omega}$ introduces detection noise arising from the inherent randomness of quantum measurements. However, this is also true for standard quantum state tomography protocols (see numerical results presented in \cite{eliska26}). A similar analysis was recently carried out for the classification of quantum correlations using two-copy collective measurements \cite{Soubusta2025:130911}. We have tested the robustness of our adaptive models against this type of noise by incorporating Poissonian randomness corresponding to a finite number $N$ of states $\hat{\Omega}$ in every iteration. Specifically, we tested the best-performing adaptive models (\emph{last} type strategy) with $n=5$ and $10$ for two-qubit states, and $n=6$ and $18$ for qubit-qutrit states. The results are summarized in Table~\ref{tab:noise}. These results indicate that the models are quite robust against detection noise for state counts of at least $N=10^5$. In these cases, the mean $\mathcal{L}^{(1)}$ errors do not exceed the noiseless baseline by more than $0.6$~\%. These results are comparable to those observed in quantum state tomography protocols, where limiting the sample size to $N=10^5$ copies of the measured state yielded a similar deviation in estimation precision \cite{eliska26}, while tomography protocols require a larger number of measurement configurations compared to collective measurements.
\begin{table}
  \caption{\label{tab:noise}
  \klepii{
  Mean values of the $\mathcal{L}^{(1)}$ error as a function of the finite number $N$ of instances of $\hat{\Omega}$ undergoing measurement in each iteration. The last row contains the $\mathcal{L}^{(1)}$ error baselines achieved without introducing detection noise ($N\rightarrow\infty$).}
  }
\begin{ruledtabular} 
 \begin{tabular}{ccccc}
    & \multicolumn{2}{c}{qubit-qubit} &  \multicolumn{2}{c}{qubit-qutrit}\\
   $N$ & $n=5$ & $n=10$ & $n=6$ & $n=18$\\
  \hline
   $10^3$ & 0.1115 & 0.0477 & 0.0913 & 0.0804 \\
   $10^4$ & 0.0438 & 0.0179 & 0.0513 & 0.0298 \\
   $10^5$ & 0.0203 & 0.0105 & 0.0440 & 0.0180 \\
   $10^6$ & 0.0154 & 0.0091 & 0.0430 & 0.0163 \\
   $\infty$ & 0.0145 & 0.0068 & 0.0427 & 0.0161 \\ 
 \end{tabular}
\end{ruledtabular}
\end{table}
}

%====================================
\section{Conclusions}

In this study, we investigated the performance of adaptive collective measurement approach to entanglement quantification in two different quantum systems: two-qubit and qubit–qutrit states. We showed that both the \emph{greedy} and \emph{last} adaptive strategies outperformed their non-adaptive counterparts, as expressed in terms of $\mathcal{L}^{(1)}$ errors and the coefficients of determination $R^2$. The superiority of adaptive methods is especially pronounced when dealing with a limited number of measurements (typically 5 and 12 for the two-qubit and qubit–qutrit systems, respectively). These results suggest that adaptive collective measurements constitute a viable and practical method for entanglement quantification in future quantum communication networks, where resource consumption in terms of measurement configurations is a concern. Note that the two-copy collective measurements discussed in this paper are experimentally quite feasible in communication devices based on entanglement swapping, such as quantum relays.

As expected, the \emph{last} approach provides somewhat better results, while the \emph{greedy} approach proved easier to train. The best-performing models were able to achieve $\mathcal{L}^{(1)}$ errors below 1.5\,\% with only 5 iterations (measurement configurations) in the case of the two-qubit states and below 2\,\% with 12 iterations in the case of the qubit–qutrit states. Within the non-adaptive strategies, we did not observe any significant difference between the \emph{greedy} and \emph{last} approaches. In general, the performance obtained meets the requirements for sufficiently small errors, especially considering that other sources of imperfections in typical implementations are on the order of 1 to 2\,\% \cite{Cernoch:PRA.97.042305,jirakova:cuncurence}. \klep{Investigation into the state-dependence of our method reveals that while the models struggle with some partially entangled states, they perform quite well on $95\%$ of the sampled states (see numerical values in Tab. \ref{tab:percentiles}). Note that the models perform much better on highly entangled states which are expected to occur in entanglement-based quantum commmunications.}

\klepii{The robustness of the models against detection noise was also evaluated, and the observations suggest that the models operate without a significant loss in performance provided that the number of instances of the measured state $\hat\Omega$ reaches at least $N=10^5$ in each iteration.}

\klep{
In addition to its potential for practical applications, the combination of collective and adaptive measurements provides opportunities for the future advancement of quantum entanglement quantification. While the idea was tested on relatively small quantum systems in this paper, it can stimulate further research to develop practical witnesses for higher-dimensional states or to investigate novel machine-designed entanglement witnesses. There are several challenges related to the transition to high-dimensional states. Namely, negativity stops being a well-behaved entanglement measure (some states are entangled despite having zero negativity). One would need to find a replacement measure to be used for training the models. Additionally, it is not presently clear which high-dimensional state would be a useful replacement for the Bell singlet state used as the non-local projector. The use of more than one non-local projector combined with sets of local projectors can have a detrimental effect on the scalability of the method. An interesting approach to mitigate this scalability problem is the decomposition of the high-dimensional state into pairs of qubit states that are tested independently \cite{7swz-fdw9}. Alternatively, the framework developed herein can be reconfigured to perform a machine search for a suitable set of high-dimensional non-local projectors to replace the Bell singlet state. Presently, in the case of qubit-qutrit systems, the non-local projection is carried out on the qubit part, while the qutrits are subjected to local projections. Even for the qubit-qutrit system, designing a single suitable qutrit-qutrit non-local projector would decrease the number of measurements needed, as one would then use the smaller set of qubit local projectors. The search for higher-dimensional non-local projectors, however, goes beyond the scope of this paper.
}
%====================================
%\section*{Acknowledgments}

\begin{acknowledgments}
The authors thank CESNET for data management services and J. Cimrman for inspiration. The authors M.Z., J.S. and K.L. acknowledge support by the project of the Czech Science Foundation No. 25-17253S; A.Č.  acknowledges support from the Ministry of Education of the Czech Republic within the OP-JAK project SENDISO no. CZ.02.01.01/00/22\_008/0004596.
\end{acknowledgments}

%\bibliographystyle{apsrev4-2}
%\bibliography{Adaptive}{}
%\setlength{\bibitemsep}{.2\baselineskip plus .05\baselineskip minus .05\baselineskip}
%\addcontentsline{toc}{chapter}{References}
%\begin{thebibliography}{99}
%\end{thebibliography}


\end{document}